\documentclass[onecolumn,showpacs,preprintnumbers,amsmath,amssymb]{revtex4}

\usepackage{graphicx}
\usepackage{dcolumn}
\usepackage{bm}
\usepackage{amsmath, amsthm, amssymb} 
\usepackage{graphicx}
\usepackage{dcolumn}
\usepackage{bm}

\newtheorem{proposition}{Proposition}
\newtheorem{definition}{Definition}

\newcommand{\bra}[1]{\langle #1|}
\newcommand{\ket}[1]{|#1\rangle}

\begin{document}


\title{Witnessed Entanglement}

\author{Fernando G. S. L. Brand\~ao}
\email{fgslb@ufmg.br}
\affiliation{Universidade Federal
de Minas Gerais - Departamento  de
F\'{\i}sica\\
Caixa Postal 702 - Belo Horizonte - MG -  Brazil - 30.123-970}
\author{Reinaldo O. Vianna}\email{reinaldo@fisica.ufmg.br}
\affiliation{Departamento de F\'{\i}sica - CP 702 - Universidade
Federal de Minas Gerais - 30123-970 - Belo Horizonte - MG -
Brazil}

\date{\today}

\begin{abstract}
We present a new measure of entanglement for mixed states. It can be approximately computable for every state and can be used to quantify all different types of
multipartite entanglement. We show that it satisfies the some of the usual
properties of a good entanglement quantifier and derive relations
between it and other entanglement measures.
\end{abstract}

\maketitle

\section{Introduction}

The quantification of entanglement in a multipartite quantum system is
one of the most important and challenging topics of quantum
information theory. The first attempt of measuring entanglement
was based on the violation of the so called Bell's inequalities
\cite{1}. It was thought that the amount of non-classical
correlations in a bipartite quantum system was intrinsically
related to the level of violation of these family of inequalities.
Nonetheless, the generalization of such inequalities to
multipartite states is not possible in general. Furthermore, it
was already shown that, even for the bipartite case, the CHSH
(Clauser-Horne-Shimony-Holt) form of Bell's inequalities does not
constitute a good measure of quantum correlations in the sense
that there are states which do not violate it, but can be purified
by local operations and classical communication (LOCC) to a state
which does \cite{3}. Therefore, different measures of entanglement
were introduced based on the concept of asymptotic distillability
\cite{4}.

For pure states, the von Neumann entropy \cite{5} of the reduced
density matrix $Tr_{B}(\ket{\psi}_{AB} {}_{AB}\bra{\psi})$ is
exactly the number of Bell states which can be asymptotically
distilled by LOCC from $\ket{\psi}_{AB}$. It satisfies all desired
properties (normalization, convexity, local unitary operations
invariance, etc...) of an entanglement measure [6]. Several
attempts were made in order to generalize this measure (the
Entropy Entanglement - $E_{E}$) for mixed sates. The most
successful are the entanglement cost ($E_{C}$), the entanglement
of formation ($E_{F}$), the relative entropy of entanglement
($E_{RE}$) and the distillable entanglement $E_{D}$ \cite{7}. They
all coincide with the entropy of entanglement in the pure states
set, showing their relationship with asymptotic distillability
properties, and satisfies most of the desirable properties.
However, none of these have an operational expression, i.e., they
cannot be calculated for arbitrary states. Furthermore, they are
defined only for bipartite states, not expressing the interesting
properties of multipartite systems \footnote{With the exeption of the relative entropy of entanglement, which is well defined also in multipartite states.} \cite{8}.

Other measures of entanglement for mixed states, not related with
$E_{E}$, have also been proposed. The robustness of entanglement
($R(\rho||{\cal S})$) quantifies the minimal amount of mixing with
separable states needed to destroy the original entanglement
presented in $\rho$ \cite{9}. It follows easily from its
definition that it can be used to quantify multipartite
entanglement. However, it is calculable only for a few specialized
bipartite density operators. The negativity $({\cal N}(\rho))$ is
the only example of a computable quantifier \cite{10}.
Nonetheless, since it is based on the negativity of the
eigenvalues of the partial transpose matrix $(\rho^{T_{A}})$ of
the system, it does not quantify the entanglement of
positive-partial-transpose (PPT) states. Moreover, it is defined
only for bipartite correlations.

In this paper we introduce a new computable measure for
entanglement, which quantifies all different kinds of multipartite
entanglement. Its definition is based on the concept of
entanglement witness. According to \cite{11}, an $n$-partite density
operator $\rho_{1...n}\in {\cal B}(H_{1} \otimes ... \otimes
H_{n})$(the space of bounded operators acting on the Hilbert space
$H_{1} \otimes ... \otimes H_{n}$) is non-separable iff there
exists a self-adjoint operator $W \in {\cal B}(H_{1} \otimes ...
\otimes H_{n})$ which detects its entanglement, i.e., such that
$Tr(W \rho_{1...n}) < 0$ and $Tr(W\sigma_{1...n}) \geq 0$ for all
$\sigma_{1...n}$ separable. This condition follows from the fact
that the set of separable states is convex and closed in ${\cal
B}(H_{1} \otimes ... \otimes H_{n})$. Therefore, as a conclusion
of the Hahn-Banach theorem, for all entangled states there is a
linear functional which separates it from this set. It will be
necessary to consider only normalized entanglement witnesses such
that $Tr(W) = 1 $.

\section{Definition}

\begin{definition}
A Hermitian operator $W_{\rho_{1...n}} \in {\cal B}(H_{1} \otimes
... \otimes H_{n})$ is an optimal EW (OEW) for the density
operator $\rho_{1...n}$ if
\begin{equation}
Tr(W_{\rho_{1...n}}\rho_{1...n}) \leq Tr(W\rho_{1...n})
\end{equation}
for every EW $W$.
\end{definition}

Although the above definition is different from the one introduced
in \cite{12}, the optimal EWs of both criteria are equal.

It is important to stress the close relation between Bell's
inequalities and entanglement witnesses. Since an EW is a
Hermitian operator, it is associated with a possible experiment
whose expectation value in all separable states is positive.
Therefore, one may consider witness operators as generalized
Bell's inequalities, where the violation of the latter is replaced
by a negative expectation value. Of course Bell's inequalities have  
an important role in the refutation of hidden loval variables theories  
for quantum mechanics, not shared by EWs. Nonetheless, in what concerns 
the detection of entanglement, entanglement witnesses can be considered just
generalized Bell's inequalities, with the locality condition of the later relaxed. Thus, the optimal entanglement witness of a state $\rho_{1...n}$ can be identified with the
experiment which most detects its entanglement, i.e., which has
the lowest possible expectation value. We will show that the
result of this \textit{optimal} experiment might be used to
quantify entanglement.

\begin{definition}
The witnessed entanglement $E_{W}(\rho_{1...n})$ of a multipartite
state $\rho_{1...n} \in {\cal B}(H_{1} \otimes ... \otimes H_{n})$
is given by
\begin{equation}
E_{W}(\rho_{1...n}) = \max {\cal f}0, - Tr(W_{\rho_{1...n}}\rho_{1...n}){\cal g}
\end{equation}
\end{definition}

The lack of an operational procedure to calculate entanglement
measures in general is ultimately related to the complexity of
distinguishing entangled from separable mixed states, which was
shown to be NP-HARD \cite{13}. Since an operational measure would
also be a necessary and sufficient test for separability, we
should not expect to find one. Nevertheless, a procedure to calculate 
OEW for all states with arbitrary probability and precision was recently 
introduced \cite{14}. Although the optimization of entanglement witnesses 
is also NP-HARD \cite{15}, the optimization of approximate EW, operators 
which are positive for almost all separable states, can be realized very 
efficiently with a chosen probability. As the amount of violation (the
percentage of negative expectation values over the separable
states set) can also be chosen, it is possible to reach the exact
OEW with any desired precision. Thus, the witnessed entanglement
(WE) introduced earlier can be approximately calculated for all
mixed states.

\section{Basic Properties}

Besides having an interesting physical interpretation and an efficient numerical method of computation, the witnessed entanglement fulfills most of the usual requirements of a
good entanglement measure \cite{16}. These properties can be summarized as:

(i) E($\sigma$) = 0 if, and only if, $\sigma$ is separable.
\vspace{0.2 cm}

(ii) Local unitary operations leave $E(\sigma)$ invariant, i.e,
\begin{equation}
E(\sigma) = E(U_{1}^{\cal y} \otimes ... \otimes U_{n}^{\cal
y}\sigma U_{1} \otimes ... \otimes U_{n}).
\end{equation}
\vspace{0.2 cm}

(iii) (Monotonicity under LOCC) Entanglement cannot increase under 
local operations and classical communication (LOCC) protocols. 
\begin{equation}
E(\Lambda(\sigma)) \leq E(\sigma),
\end{equation}
where $\Lambda$ is a superoperator implementable by LOCC.
\vspace{0.2 cm}

(iv) (Continuity) For every $\epsilon \geq 0$ and density operators $\rho$ and $\sigma$, there is a real number $C \geq 0$ such as
\begin{equation}
||\rho - \sigma||_{L} \leq \epsilon \Rightarrow |E(\rho) - E(\sigma)| \leq C(L)\epsilon,
\end{equation} 
where $||$ $||_{L}$ is any norm defined for finite dimensional systems. Note that $C$ depends of the chosen norm.
\vspace{0.2 cm}

(v) (Convexity) Mixing of states does not increase entanglement.
\begin{equation}
E(\lambda \rho + (1 - \lambda)\sigma) \leq \lambda E(\rho) + (1 -
\lambda)E(\sigma),
\end{equation}
for all $\rho$, $\sigma$ and $0 \leq \lambda \leq 1$.

\begin{proposition}
The Witnessed Entanglement satisfies properties (i), (ii),
(iv) and (v).
\end{proposition}

\begin{proof}

\vspace{0.2 cm}
(i): From the definition of entanglement witnesses we have that
$Tr(W_{\sigma}\sigma) \geq 0$ for all separable states $\sigma$.
Therefore, 
\begin{equation}
E_{W}(\sigma) = \max{\cal f} 0,  - Tr(W_{\rho_{1...n}}\rho_{1...n}){\cal g}
= 0.
\end{equation}
Conversely, as there exists an EW for every entangled state, $E_{W}$
is positive for any non-separable density operator.

\vspace{0.2 cm}
(ii): We will prove by absurd that $E_{W}$ is invariant under
local unitary operations. Suppose that $E_{W}(\sigma) >
E_{W}(\rho)$, where 
\begin{equation}
\sigma = U_{1}^{\cal y} \otimes ... \otimes
U_{n}^{\cal y}\rho U_{1} \otimes ... \otimes U_{n}.
\end{equation}
Then,
\begin{eqnarray}
E_{W}(\sigma) = {\cal f}- Tr(W_{\sigma}\sigma), 0 {\cal g} = \max {\cal f}- Tr(W_{\sigma}U_{1}^{\cal y}
\otimes ... \otimes U_{n}^{\cal y}\rho U_{1} \otimes ... \otimes
U_{n}), 0 {\cal g} \\ \nonumber
= \max {\cal f} - Tr(W \rho), 0 {\cal g} > \max {\cal f}- Tr(W_{\rho}\rho), 0 {\cal g}
\end{eqnarray}
where $W = U_{1} \otimes ... \otimes U_{n}W_{\sigma}U_{1}^{\cal y} \otimes
... \otimes U_{n}^{\cal y}$ is an entanglement witness. But equation (9)
contradicts the fact that $W_{\rho}$ is the OEW for $\rho$. If we
suppose that $E_{W}(\sigma) < E_{W}(\rho)$, the same argument can
be applied to $\rho = U_{1} \otimes ... \otimes U_{n}\sigma
U_{1}^{\cal y} \otimes ... \otimes U_{n}^{\cal y}$. Therefore,
$E_{W}(\rho) = E_{W}(\sigma)$ must hold.

\vspace{0.2 cm}
(iv): Let $W_{\rho}$ and $W_{\sigma}$ be optimal entanglement witnesses for $\rho$ and $\sigma$, respectively. If either $\rho$ or $\sigma$ is separable, the result follows trivially. We then assume that   
\begin{equation}
Tr(W_{\rho}\rho) < 0, \hspace{0.2 cm}  Tr(W_{\sigma}\sigma) < 0. \hspace{0.4 cm} (a)
\end{equation}
As the intersection of the Ews set with the set of Hermitian matrices with unity trace is compact, we have that for some norm 
$||$ $||_{L'}$,
\begin{equation}
\max_{W \in {\cal M}}||W||_{L'} = D, \hspace{0.4 cm}(b)
\end{equation}
where $D \geq 0$ is a real number. Every norm of finite demensional operators are equivalent, i.e., for every finite dimensional operator $A$ and any two chosen norms $||$ $||_{L'}$ and $||$ $||_{L}$, there always exists rela numbers $n$ and $m$ such that
\begin{equation}   
m||A||_{L} \leq ||A||_{L'} \leq n||A||_{L} \hspace{0.4 cm}(c).
\end{equation}
Thus, one sees that equation (14) is valid for every norm.

We can assume without loss of generality that $E(\rho) \geq E(\sigma)$ (d). Hence,
\begin{equation} 
|E(\sigma) - E(\rho)| = |Tr(W_{\rho}\rho) - Tr(W_{\sigma}\sigma)| = |Tr[W_{\rho}(\rho - \sigma)] + Tr(W_{\rho}\sigma) - Tr(W_{\sigma}\sigma)|.
\end{equation}
Since $W_{\sigma}$ is a OEW for $\sigma$, $Tr(W_{\rho}\sigma) - Tr(W_{\sigma}\sigma) \geq 0$. From (d) we also have $|Tr[W_{\rho}(\rho - \sigma)]| \geq |Tr[(W_{\rho} - W_{\sigma})\sigma]|$. Therefore, 
\begin{equation} 
|E(\sigma) - E(\rho)| = |Tr[W_{\rho}(\rho - \sigma)] + Tr(W_{\rho}\sigma) - Tr(W_{\sigma}\sigma)| \leq |Tr[W_{\rho}(\rho - \sigma)]| \leq ||W_{\rho}||_{HS}||\rho - \sigma||_{HS},
\end{equation}
where we have used the Cauchy-Schwartz inequality for the Hilbert-Schmidt inner product. From Eq. (17) and conditions (b) and (c) we then have that for every norm $||$ $||_{L}$, 
\begin{equation}
|E(\sigma) - E(\rho)| \leq ||W_{\rho}||_{HS}||\rho - \sigma||_{HS} \leq a||W_{\rho}||_{L}||\rho - \sigma||_{L} \leq C\epsilon,
\end{equation}
for some real numbers $a, C \geq 0$.

(v):
The convexity follows straightforwardly from definition 2 and the concavity and convexity of the $\max$ and $\min$ functions, respectively.
\end{proof}

Although $E_W$ is not a monotone under general LOCC maps, we can state the following:
\begin{proposition}
For every unital LOCC map ${\cal \Lambda}$,
\begin{equation}
E_W(\Lambda(\rho)) \leq E_W(\rho)
\end{equation}
\end{proposition}
\vspace{0.2 cm}

\begin{proof}
It suffices to show that $E_{W}$ is not increasing under
separable operations. Let $\Lambda$ be a uninal trace-preserving separable quantum superoperator. Then, 
\begin{equation}
Tr(W_{\rho'}\Lambda(\rho)) = Tr(\Lambda^{\cal y}(W_{\rho})\rho),
\end{equation}
where $\Lambda^{\cal y}$ is the dual map of $\Lambda$, which is unital, since $\Lambda$ is trace preserving, and separable, and $W_{\rho'}$ is the optimal EW for $\rho' = \Lambda(\rho)$., $\Lambda^{\cal y}(W_{\rho})$ is also an EW. Indeed, $Tr(\Lambda^{\cal y}(W_{\rho})\sigma) = Tr(W\Lambda(\sigma)) \geq$ for all $\sigma \in {\cal S}$, since $W$ is an EW and $\Lambda(\sigma) \in {\cal S}$. Now we have to prove that $Tr(\Lambda^{\cal y}(W_{\rho})) = 1$. Using that the channel is unital $T$, i.e. $T(I) = I$, 
\begin{equation}
Tr(\Lambda^{\cal y}(W_{\rho})) = Tr(W_{\rho}) = 1.
\end{equation}
\end{proof}

\section{Multipartite Entanglement}

A $n$-partite density operator $\rho_{1...n}$ is a $m$-separable state
if it is possible to find a decomposition for it such that in each
pure state term at most m parties are entangled among each other,
but not with any member of the other group of $n - m$ parties.
Since the subspace of m-separable density operators is convex and
closed, it is also possible to apply the Hahn-Banach theorem to it
and establish the concept of entanglement witness to
$(m+1)$-partite entanglement. In order to do that, consider the
index set $P = {\cal f}1, 2, ..., n{\cal g}$. Let $S_{i}$ be a
subset of P which has at most m elements. Then $W$ is an
$(m+1)$-partite entanglement witness if:
\begin{center}
\begin{equation}
\begin{array}{c}
 _{P^{m}_{v}}\bra{\psi}\otimes ... \otimes \hspace{0.07 
cm}_{P^{m}_{1}}\bra{\psi}W\ket{\psi}_{P^{m}_{1}} \otimes ... \otimes 
\ket{\psi}_{P^{m}_{v}} \geq 0 \\ \\
\forall \hspace{0.2 cm} P^{m}_{1}, ..., P^{m}_{v} \hspace{0.2 cm} $such 
that$ \\ \\
\bigcup_{k=1}^{v}P^{m}_{k} = P \hspace{0.2 cm} $and$ \hspace{0.2 cm} 
P^{m}_{k} \bigcap P^{m}_{l} = {\cal f}{\cal g} 
\end{array}
\end{equation}
\end{center}

Equation $(19)$ assures that the operator $W$ is positive for all
m-separable states. Thus, it is also possible to determine optimal
$(m+1)$-partite entanglement witnesses for every $(m+1)$-partite
entangled state and establish an order in this set with $E_{W}$.

\section{Connections with Other Entanglement Measures}

We have considered so far OEW of the most general form. We will
show that it is possible to construct another measure of
entanglement for non-positive-partial-transpose-states (NPPTES),
reducing the search space of EW. Witnesses operators can be
classified as decomposable (d-EW) or not decomposable (nd
-EW)\cite{12}. In the case of bipartite entanglement, a Hermitian
operator $W$ is a $d-EW$ if it can be written in the form of
\begin{equation}
W = pP + (1 - p)Q^{T_{A}}, \hspace{0.3 cm} 0 \leq p \leq 1
\end{equation}
where P and Q are positive operators. It is clear from this
definition that these EWs cannot detect PPTES. A $nd-EW$ is a
Hermitian operator which cannot be written as $(20)$. It was
already shown that $W$ is a $nd-EW$ if, and only if, it detects
PPTES [12]. If we restrict our search of optimal EW to $d-EW$, we
will still have a measure of entanglement, the decomposable
witnessed entanglement ($E_{d-W}$), which satisfies the same
properties of $E_{W}$, despite the fact that it will be null for
both separable and PPTES. The advantage of this new measure is
that it can be calculated in a straightforward manner.
\begin{proposition}
$E_{d-W}(\rho_{AB}) = log(d^{D/d})$
$|min(\lambda_{min}(\rho_{AB}^{T_{A}}), 0)|$, where
$\lambda_{min}(A)$ stands for the minimum eigenvalue of A.
\end{proposition}
\begin{proof} 
If $\rho_{AB}$ is separable or PPT, then
$Tr(d-W_{\rho_{AB}}\rho_{AB})$ will be non-negative and $E_{d-W}$
will be null. If $\rho_{AB}$ is NPPT, since $Tr(XY^{T_{A}}) =
Tr(X^{T_{A}}Y)$, it is easy to see that the optimal d-EW will be
$P^{T_{A}}$, where $P$ is the projector in the subspace of the
minimum eigenvalue of $\rho_{AB}^{T_{A}}$. 
\end{proof}

The negativity ($\cal{N}$) is another example of an entanglement
measure which is both computable and null for PPTES \cite{10}. It
is well known that, for systems of two qubits, $\cal{N}$ is equal
to twice the absolute value of the negative eigenvalue of the
partial transpose of the state. Thus, in this dimension, it
coincides with $E_{d-W}$. The decomposable witnessed entropy of
entanglement is particularly interesting for 2 x 2 and 2 x 3
systems. Since there are no PPT entangled states in these cases,
the optimal $d-EW$ is the exact OEW and $E_{W}$ and $E_{d-W}$ are
equal.
\begin{proposition}
$E_{W}(\rho) = E_{d-W}(\rho)$ and $E_{W}(\rho) = E_{d-W}(\rho) =
{\cal N}(\rho)$ for all 2 x 3 and 2 x 2 states, respectively.
\end{proposition}

We have seen that, up to a dimension normalization factor,
$E_{W}(\rho)$ can be identified with the result of an experiment
which has the lowest possible expectation value  in $\rho$, while
it  has non-negative expectation value in all separable states.
$E_{W}(\rho)$ is also related to the amount of mixing with the
maximally random state $(I/D)$ needed to destroy the original
entanglement presented in $\rho$. This quantity, introduced in
[9], is called random robustness ($R(\rho||I/D)$) and is given by
the minimum $s$ for which $\rho(s) = \frac{1}{1 + s}(\rho + sI/D)
$ is separable.
\begin{proposition}
$E_{W}(\rho) = R(\rho||I/D)/D$, for every multipartite density
operator $\rho$.
\end{proposition}
\begin{proof}
Since $s' = R(\rho||I/D)$ is the minimum value
for which $\rho(s)$ is separable, $Tr(W_{\rho}\rho(s')) =
\frac{1}{1 + s'}(Tr(W_{\rho}\rho) + s'/D) = 0$. Therefore,
$R(\rho||I/D) = - DTr(W_{\rho}\rho)$. 
\end{proof}

It was proved in \cite{9} that, for bipartite pure states,
$R(\rho||I/D)$ is given by the product of  the two biggest Schmidt
coefficients with the dimension of the composite Hilbert space .
\begin{proposition}
Given a pure bipartite state $\ket{\Psi}$ with decreasing ordered
Schmidt coefficients $a_{i}$, $E_{W}(\ket{\Psi}\bra{\Psi}) =
a_{1}a_{2}$.
\end{proposition}

Since $R(\rho||I/D)$ is an upper bound for the robustness
$R(\rho||{\cal S})$ \cite{9}, it is possible to use the results
presented in \cite{10} and establish the following order between
different measures: 
\begin{equation}
E_{W}(\rho) \geq
R(\rho||{\cal S}) \geq {\cal N}(\rho) \geq E_{D}(\rho).
\end{equation}

\section{Example}

As an example, consider the following state of three qubits: $\rho_{ABC} = (1
- p)\ket{\psi_{W}}\bra{\psi_{W}} +
p\ket{\psi_{GHZ}}\bra{\psi_{GHZ}}$, where $\ket{\psi_{GHZ}} =
\frac{1}{\sqrt{2}}(\ket{000} + \ket{111})$ and $\ket{\psi_{W}} =
\frac{1}{\sqrt{3}}(\ket{100} + \ket{010} + \ket{001})$. We have
calculated $E_{W}$ for genuine tripartite entanglement, bipartite
entanglement of the possible partitions (AB-C, A-BC and AC-B), and
bipartite entanglement of the reduced density matrices (A-B, A-C
and B-C). The results are plotted in Fig. 1. Since the state is
symmetric with respect to exchange of the particles, the amount of
bipartite entanglement in the three partitions and in the three
reduced density matrices are the same. The value of $E_{d-W}$,
which was also calculated for the three partitions, and $E_{W}$
are the same, showing that it is not possible to reach a bound
entangled state mixing the identity with this family of states. Note in addition that the the witnessed entanglement gives the same order than the relative entropy of entanglement \cite{19} and the geometric measure of entanglement \cite{20} for the W and GHZ states.
\begin{figure}
\includegraphics[scale=0.7]{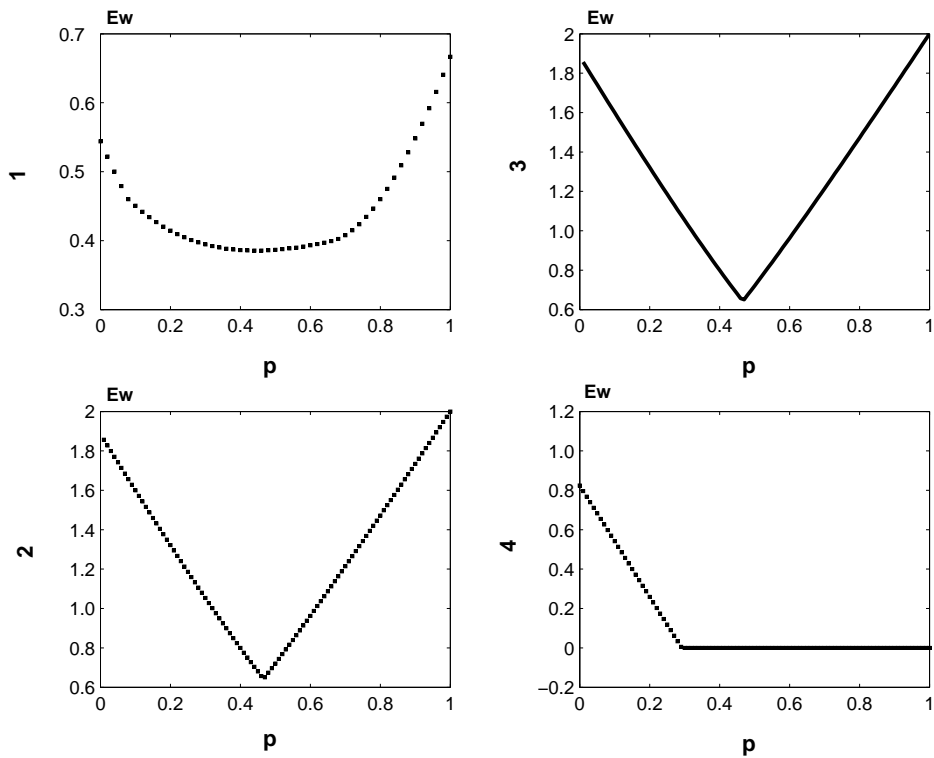}
\caption{ $E_{W}$ as a function of $p$ for the tripartite
entanglement (1), for the bipartite entanglement in any of the
three partitions (2), for the bipartite entanglement in any of the
three reduced density matrices (4), and $E_{d-W}$ as a function of
$p$ for the bipartite entanglement in any of the three partitions
(3).}
\end{figure}

\section{Conclusion}

In summary,  we have introduced a new measure of entanglement
which is operational and can be used to quantify all different
types of multipartite entanglement. It has two interesting
physical interpretations and is connected to several other
different measures. 

\begin{acknowledgments}

The authors wish to thank Marcelo O. Terra Cunha for his precious
comments and revisions. Partial financial support from the
Brazilian agencies CNPq, Institutos do Mil\^enio-Informa\c{c}\~ao
Qu\^antica and FAPEMIG.

\end{acknowledgments}

\end{document}